# Rotational Quasi-Periodicities and the Sun – Heliosphere Connection

J. K. Lawrence[1], A. C. Cadavid[1], and A. Ruzmaikin[2]

[1]Department of Physics and Astronomy, California State University, Northridge, 18111 Nordhoff Street, Northridge, California 91330-8268, USA

(e-mail: *john.lawrence@csun.edu*, *ana.cadavid@csun.edu* )

[2]Jet Propulsion Laboratory, California Institute of Technology, 4800 Oak Grove Drive, Pasadena, California 91109, USA

(e-mail: *alexander.ruzmaikin@jpl.nasa.gov* )

**Abstract.** Mutual quasi-periodicities near the solar-rotation period appear in time series based on the Earth's magnetic field, the interplanetary magnetic field, and signed solar-magnetic fields. Dominant among these is one at 27.03 ± 0.02 days that has been highlighted by Neugebauer, *et al*. 2000, *J. Geophys. Res.*, **105**, 2315. Extension of their study in time and to different data reveals decadal epochs during which the ≈ 27.0 day, a ≈ 28.3 day, or other quasi-periods dominate the signal. Space-time eigenvalue analyses of time series in 30 solar latitude bands, based on synoptic maps of unsigned photospheric fields, lead to two maximally independent modes that account for almost 30% of the data variance. One mode spans 45º of latitude in the northern hemisphere and the other one in the southern. The modes rotate around the Sun rigidly, not differentially, suggesting connection with the subsurface dynamo. Spectral analyses yield familiar dominant quasi periods 27.04 ± 0.03 days in the North and at 28.24 ± 0.03 days in the South. These are replaced during cycle 23 by one at 26.45 ± 0.03 days in the North. The modes show no tendency for preferred longitudes separated by ≈ 180º.

## 1. Introduction

Quasi-periodicities, that is periodicities that come and go in time, can reveal connections among related physical systems. In the present paper we explore two sets of systems on time scales near the Sun's Carrington rotation period. The first consists of three sets of time series related to the polarity of the interplanetary magnetic field (IMF). One is based on a reconstruction of the IMF polarity from terrestrial magnetometer measurements of the geomagnetic polar field. Another is from *in situ* spacecraft measurements of the radial component of the IMF, and yet another is an estimate of the equatorial coronal source surface field derived from ground-based observations of solar photospheric fields. The second set of systems comprises time series of unsigned photospheric magnetic fields in 30 adjacent latitude bands from pole to pole on the Sun. We will see that these latitude bands tend to undergo collective fluctuations extending across many latitudes. These collective modes appear, on the one hand, to be related to the subsurface solar dynamo, and, on the other hand, also to be related to fluctuations in the IMF and polar geomagnetic field. In all cases our attention is confined to fluctuations in the data on time scales near the solar rotation period ≈27 days.

Studies of this kind are not novel, and a sizeable literature with numerous contributors exists. We will touch on many of these below. See Henney and Harvey (2002), Bai (2003), and Knaack, Stenflo and Berdyugina (2005) and references therein. Nevertheless, further investigation can still improve understanding. Some of the previous work was performed on early data and one or two additional solar cycles have provided new data. In other cases, we are in a position to apply advanced methods to data and to see them from a different perspective. We will highlight research most closely related to the present study, and we will make explicit comparisons to previous results in the discussion section below.

The first motivation for our work has been the result of Neugebauer, *et al*. (2000) who found evidence in the solar wind (SW) and IMF for a dominant solar rotation period at 27.03 ± 0.02 days. This result was based on data from 1963 to 1998, covering solar activity cycles 20 – 22. An obvious extension, a decade later, is to include cycle 23 as well. An additional extension, into the past, also is possible. Of prime interest is the data set of Svalgaard (1973) based on daily inferences from geomagnetic observations of IMF polarity. This data set extends from 1926 to the present, and includes cycles 16 – 23.

Another extension of this study is the inclusion of contemporaneous direct observations of solar fields. Henney and Harvey (2002) used NSO/Kitt Peak synoptic maps to identify the 27.03 day periodicity with particular solar active regions. However, Henney and Durney (2005) have cast doubt on the statistical significance of this result. In earlier work, Antonucci, Hoeksema, and Scherrer (1990) took note of dominant periodicities of 27 days and 28.5 days in the IMF polarity data (Svalgaard 1973) from 1926 to 1986. Then, using Wilcox Solar Observatory (WSO) synoptic magnetic field maps, they related these periods, at least during solar activity cycle 21, to a 26.9 day northern hemisphere magnetic pattern and a 28.1 day southern hemisphere pattern. These extended over wide latitude intervals (24º to 32º). We will arrive at similar results from cycle 21 through the end of cycle 23. First, we extend the IMF study from cycle 21 through cycles 22 and 23. Then, looking at the solar synoptic maps shows that almost 30% of the unsigned solar magnetic field appears in independent, rigidly rotating, northern and southern hemispheric patterns extending over some 40º of latitude. The southern pattern shows a strong periodicity at 28.24 ± 0.03 days, and the northern pattern shows a strong periodicity near 27.04 ± 0.03 days. These periodicities remain strong through cycles 21 and 22 but vanish in cycle 23, when a northern periodicity at 26.45 ± 0.03 days appears.

An investigation by Knaack, Stenflo, and Berdyugina (2005) of the temporal spectra of solar magnetic structures includes results consistent with ours at the short-period end of a broader examination of longer timescale phenomena. There are differences in detail, because, although they consider both the signed and unsigned fields, they do not equalize solar maxima and minima, and thus omit periodicities in the quiet-Sun intervals. These include, for example, strong periodicities around 26.5 days in the North between cycles 21 and 22 and in the South following cycle 23.

In Section 2 we introduce the data sets we employ in the present work and indicate how the data are prepared. Section 3 describes the data analysis procedures and presents the results thereof. The results are discussed and conclusions are drawn in Section 4.

## 2. Data and Data Preparation

We have gathered, in this section, descriptions of the data sets we make use of in our analyses below, together with accounts of the data preparation procedures.

### 2.1 IMF *in Situ* Spacecraft Measurements

We first extended the radial interplanetary magnetic field data set used by Neugebauer, *et al*. (2000) from the period 1963 – 1998 to the period 1963 – 2007, and thus included activity cycle 23. The particular parameter used was the radial magnetic field in the Radial-Tangential-Normal coordinate system. The data were obtained from NASA's SPDF data base at http://cohoweb.gsfc.nasa.gov/. The new data from the Coordinated Heliospheric Observations (COHO) compilation are from the *Wind* and ACE spacecraft near 1 AU.

In their work Neugebauer, *et al*. (2000) removed the effect of the 11-year polarity reversal of solar fields by reversing the sign of the IMF during alternate cycles. Although we have used slightly different dates for reversing the polarities, we have adopted the same procedure for comparison. We have reversed the polarities in this and all of our signed data sets below in the intervals 1926.06 – 1935.0, 1945.1 – 1955.3, 1965.8 – 1977.4, and 1987.3 – 1997.3.

**2.2 Inferred IMF Polarity**

A second, longer, extension of this investigation used reconstructions of the IMF polarity at the Earth that are based on terrestrial magnetometer observations of the polar magnetic field (Svalgaard, 1973). These reconstructed data give daily inferences of whether the IMF field at the Earth is directed toward or away from the Sun along the Parker spiral. The data extend from early 1926 to the present, and they were retrieved from http://www.leif.org/research. For consistency, we have applied the polarity reversal in alternate solar activity cycles as described in Section 2.1 above. Finally, these data are represented as sequences of ±1, with a small number of ambiguous cases represented by 0.

**2.3 WSO Synoptic Data**

The magnitudes of the radial interplanetary magnetic field, described in Section 2.1 above, have little or no correlation with the phase of the solar cycle. The IMF polarity data in Section 2.2, being composed of ±1 have no such correlation. However, the magnitudes of the solar synoptic data we are now concerned with are strongly correlated with the activity cycle. Therefore, for consistency with the IMF data sets, and to de-emphasize solar cycle changes in field strengths in favor of signal periodicities, the data in all WSO time series we use are divided by a running 27-day standard deviation.

*2.3.1 Coronal 2.5 Solar Radii Source Surface*

We next extend our analysis to WSO synoptic maps of 2.5 solar radii coronal source surface magnetic fields. We use the maps derived from photospheric magnetograms via potential field extrapolation that assume radial photospheric fields; see http://wso.stanford.edu . We average the values of the central six of the 30 sin(latitude) rasters, bracketing the equator, *i.e.* between latitudes ± 11º.5. The averages from all the synoptic maps are then concatenated to make a single time series with interval 27.2753 / 72 = 0.3788236 days (or 5º of Carrington longitude). The time series used extended from 1976.38 to 2007.22. Note that in the WSO synoptic charts time goes from right to left, oppositely to NSO/Kitt Peak synoptic charts. If one concatenates them in the wrong sense, it is spectacularly obvious in the analysis.

Again for consistency, we have applied the polarity reversal in alternate solar activity cycles as described in Section 2.1 above.

*2.3.2 Photospheric Synoptic Plots*

The last data set to be considered here is derived from the WSO synoptic maps of photospheric magnetic field. Again see http://wso.stanford.edu. The goal is to examine the latitudinal and longitudinal distribution of solar activity. Because we wish to study the distribution of solar activity in general, and because the details of the polarity structure of active regions complicate the analysis, we take the absolute values of the fields. In keeping with this, and to simplify analysis we study the unsigned (absolute value) fields. The unsigned synoptic maps are concatenated to make 30 time series extending from 1976.38 to 2007.22 with time step 0.3788236 days (corresponding to 5º of Carrington longitude). The 30 series cover all latitudes on the Sun from pole to pole in bands of equal sin(latitude). Because our interest is on the solar rotation and not on magnetic field strength, the 30 time series are divided by a running 27-day standard deviation to standardize the quiet-Sun and active-Sun times.

## 3. Analysis and Results

**3.1 Geomagnetic, Interplanetary, and Coronal Magnetic Field**

3*.1.1 Quasi-Periodicities in the Extended Heliospheric Time Series*

With the data preparation changes noted above, we have first calculated the spectra for fluctuations of the radial component of the IMF, using data from the COHO website at http://cohoweb.gsfc.nasa.gov/, for the same time interval studied by Neugebauer *et al*. (2000). Since there are gaps in the data, we have used the Lomb-Scargle periodogram method (Scargle, 1982). For these and all periodogram calculations we employ the AutoSignal *version* 1.0 package from Systat Software, Inc. The result is shown in Figure 1, and closely resembles their earlier result, with a dominant periodicity appearing at 27.03 days. Because our main interest is in the influence of the solar rotation and dynamo on the heliosphere and the Earth, this periodogram is restricted to periods in the neighborhood of 27 days. Numerous harmonics at shorter periods have been ignored. We have extended the Neugebauer, *et al*. (2000) investigation by lengthening the data set we use to the period 1963 – 2007 and thus including activity cycle 23. A Lomb-Scargle periodogram of the expanded IMF radial component data set also is shown in Figure 1. In comparison to the result of Neugebauer, *et al*. (2000) the addition of subsequent data has reduced the relative importance of the main spectral peak near period 27.03 days and has enhanced the importance of side peaks, especially those at 26.8 days and 27.65 days. We note that during 1999 – 2007 the only clear, though broad, spectral peaks were around 26.5 days and 27.2 days.

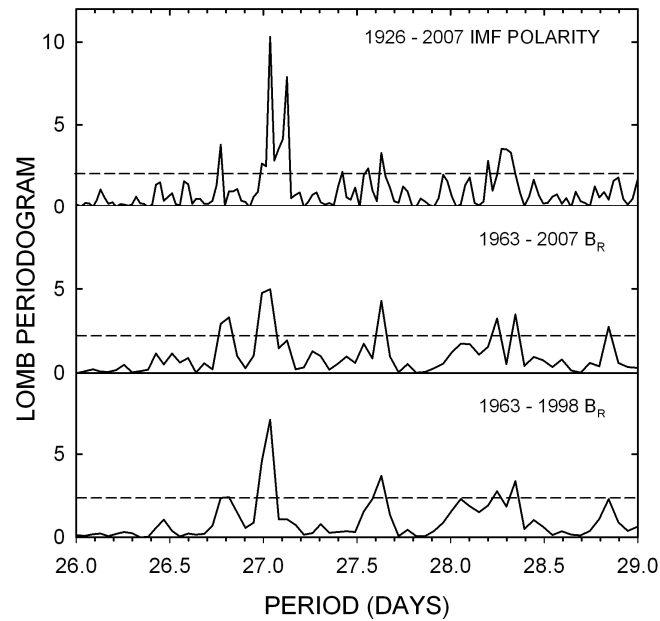

**Figure 1.** Lomb periodograms of the original 1963 – 1998 radial component of the IMF from the COHO data set (bottom), of the expanded 1963 – 2007 COHO data (middle), and of the 1926 – 2007 reconstruction of the IMF polarity from ground based magnetometer data (top). The horizontal dashed lines indicate the 99.9% significance levels of the corresponding spectra. The periodograms have been normalized to a mean value of unity.

A periodogram of the reconstructed IMF polarity data described in Section 2.2 is shown in Figure 1 as well. Because the polarity data resemble a square wave, and not a sinusoid, there are strong harmonics in the spectrum. For the reason given above, however, we did not consider them. Figure 1 shows that the spectral peak at period 27.03 days is strengthened in the longer data set. In addition, the side peaks in this spectrum generally coincide in period and amplitude. (Each periodogram is normalized to an average value of unity.) A new peak at period 27.12 days appears in the IMF polarity periodogram.

Figure 2 presents, in comparison to one another, periodograms of the WSO source surface equatorial data series described in Section 2.3.1, the daily IMF polarity data described in Section 2.2, and the COHO radial magnetic field data described in Section 2.1. All three of these are over the same 1976.38 – 2007.22 time span. A number of spectral peaks from the three, geomagnetic, IMF and solar, time series are seen to coincide. For this time interval, there are two peaks very near 27 days: one at 27.06 days, another at 26.98 days. The peak at 26.8 days is strong in all three series. Note for future reference the peaks near 26.5 days and 28.25 days.

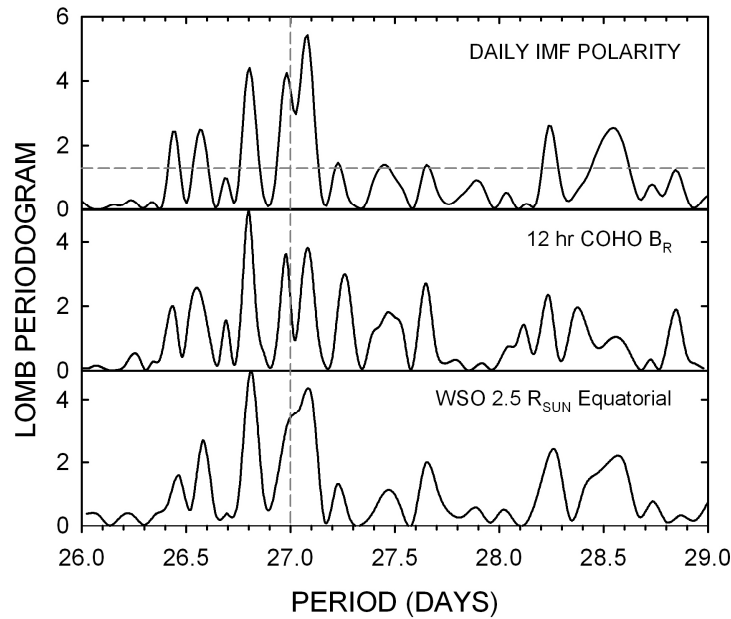

**Figure 2.** Lomb periodograms of the daily IMF polarity time series (top), the 12 hr COHO radial magnetic field series (middle), and the WSO 2.5 $B_R$ equatorial coronal field series (bottom) over the time span 1976.38 – 2007.22. The horizontal dashed line represents the most conservative of the three 99.9% confidence levels. The vertical dashed line gives a reference at period 27.00 days. These periodograms have been normalized to a mean value of unity.

The key result here is that the ≈ 27 days rotation spectra in the terrestrial, IMF and solar data resemble one another in considerable detail. Furthermore, all three spectra display the ≈ 27.03 day period of Neugebauer, *et al*. (2000).

*3.1.2 Decadal Persistence of Dominant Quasi Periodicities Predominant*

The Lomb-Scargle periodogram method fits pure sinusoidal functions $Y(t) \equiv \sin(\omega_0 t + \delta_0)$, with constant amplitude, frequency $\omega_0$ and phase $\delta_0$ to the full span of the data and computes the goodness of fit. Thus, while the periodograms in Figures 1 and 2 indicate which periodicities in the data are present, they do not indicate when each such periodicity occurs or when it dominates. Periodograms for subintervals frequently do not resemble that for the whole data set.

A technique that allows us to evaluate temporal changes of amplitude and frequency is "complex demodulation" (Bloomfield, 1976). We adopt the test function

$$Y(t) \equiv a(t) \sin[\omega_0 t + \delta(t)], \qquad (1)$$

where now the amplitude and phase are functions of time, and $\omega_0$ is an initial constant frequency selected to be near the true frequency of the signal. Then we filter the signal to eliminate those components that are not near frequency $\omega_0$ and estimate the phase $\delta(t)$. If the signal frequency is in fact equal to $\omega_0$, then the phase $\delta(t)$ will be constant in time. If the signal frequency is a constant $\neq \omega_0$ then $\delta(t)$ will be a linear function of time. More importantly for our analysis, if the signal consists of a sinusoid of frequency $\omega_0$ plus a second, weaker, sinusoid of different frequency $\omega_1$, then the phase $\delta(t)$ will fluctuate with the beat period corresponding to $\omega_0$ and $\omega_1$ around a constant value. Figure 3 shows the phase $\delta(t)$ for the daily IMF polarity with test frequencies $2\pi/\omega_0$ = 27.03 days and 28.30 days. The plot is very compressed in time, and the phase varies smoothly although it may appear discontinuous.

The striking aspect of the plot in Figure 3 is that there are epochs, on decadal time scales, during which particular periods dominate. For example, there are two time intervals, 1926 – 1950 and 1980 – 1992, where the dominant period is near 27.03 days. During the intervening time, longer periods dominate. Figure 3 also shows the phase $\delta(t)$ for test period $2\pi/\omega_0$ = 28.30 days. This period appears dominant during the time interval 1950 – 1970.

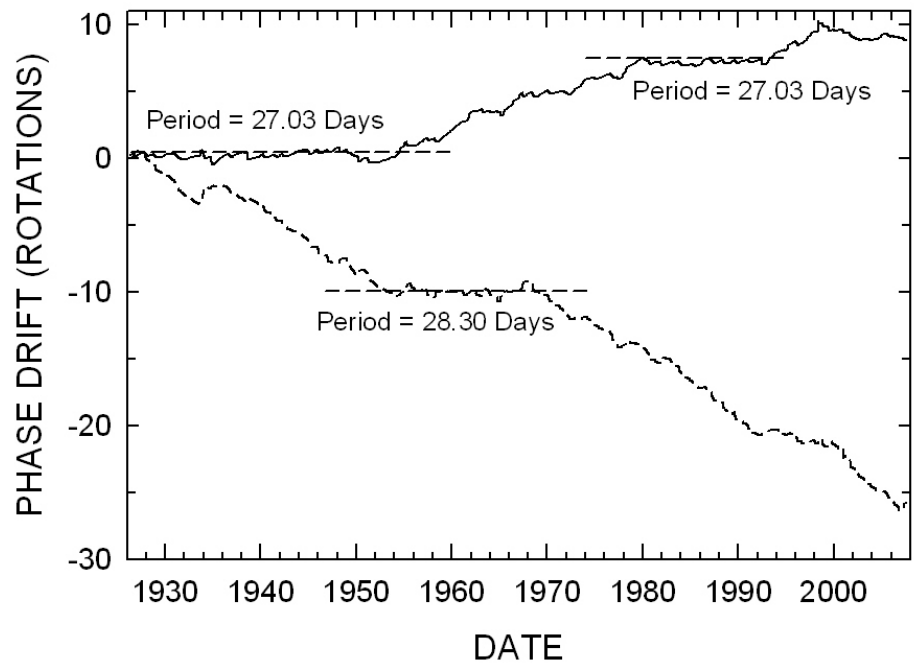

**Figure 3.** Phase function $\delta(t)$ for the daily IMF polarity data versus time for the test period $2\pi/\omega_0$ = 27.03 days (solid line) and for the test period 28.30 days Dashed line. The horizontal dashed lines indicate time intervals during which either the 27.03 day or the 28.30 day period appears to dominate the IMF polarity signal.

The complex demodulation method indicates the time intervals at which certain periodicities are dominant in the data, and this indicates decadal scale regimes of apparent dominance by one period or another. We cross-check these results by fixing the signal frequency at $\omega_0$ and instead estimating the amplitude $a(t)$ in Equation (1). The result for the two test periods $2\pi/\omega_0$ = 27.03 days and 28.30 days is shown in Figure 4. This method confirms the result of the complex demodulation. The amplitude for the period $2\pi/\omega_0$ = 27.03 days, although it is highly variable, is generally the greatest one from 1926 to 1950, from 1980 to 1992, and after 2000. Likewise, the amplitude for $2\pi/\omega_0$ = 28.30 days tends to be the largest from 1950 to 1970.

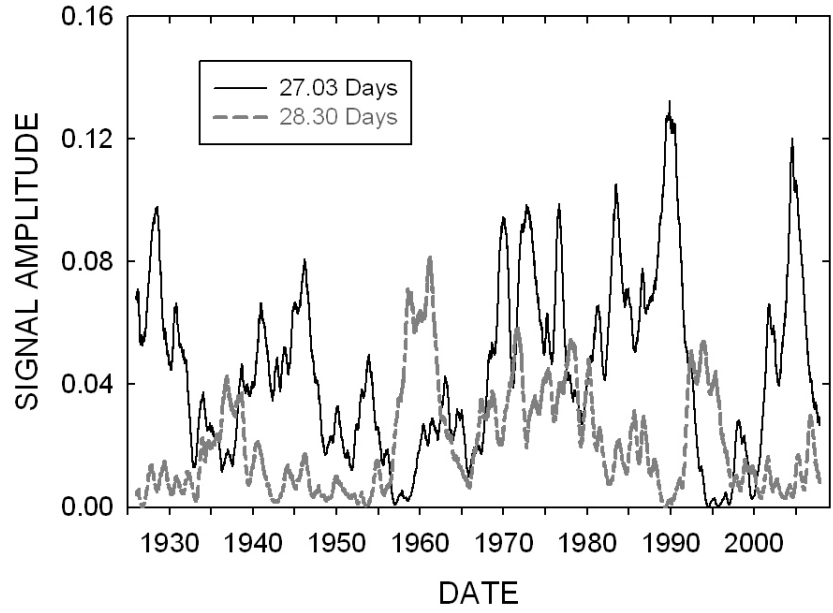

**Figure 4.** Plot of the amplitude function *a(t)* in Equation (1) for test periods: 27.03 days (solid black) and 28.30 days (dashed gray).

**3.2 Photosphere and Dynamo**

*3.2.1 Characteristic Quasi Periods of Synoptic Data*

Next we turn more directly to solar photospheric data, starting with Wilcox Solar Observatory (WSO) Carrington maps of the photospheric magnetic field. We concatenate the maps end-to-end to generate 30 time series for the unsigned field in 30 bands of sin(latitude) from the North to the South pole. The signal in each series is divided by a running 27-day standard deviation to equalize active and quiet Sun and thus to focus on the rotation. The spectra cover the interval from 1976.44 to 2007.39. Because the high-latitude data are foreshortened and also are contaminated by a strong annual signal, Lomb-Scargle periodograms were calculated only for the central 24 of the series, covering sin(latitude) = ± 0.8. The spectra for periods between 26 and 29 days are shown stacked in Figure 5. Only spectral features significant above the 99% confidence level are shown.

Significantly, the spectra in Figure 5 show the presence of strong structures that rotate rigidly across wide latitude bands. We can see such a structure at the noteworthy period 27.04 days that extends from south of the solar equator to some +35º of north latitude. Another at 26.99 days extends to the South. These, and some other spectral features, lie across the differential rotation curve calculated from day-to-day magnetic field correlations (Komm, Howard, and Harvey, 1993), but others lie completely outside. The features with periods less than 26.75 days rotate more quickly than any part of this differential rotation curve. These reveal the existence of spatially extended magnetic patterns that rotate rigidly, remain coherent over several rotation periods and are not governed by day-to-day surface flows. We next explore the nature of these patterns.

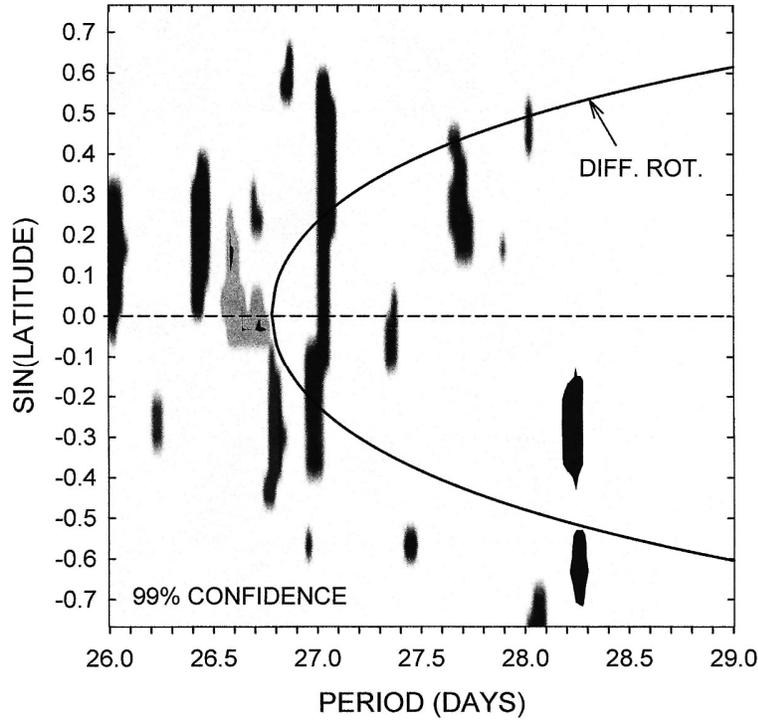

**Figure 5.** 24 Lomb-Scargle periodograms of |**B**| in WSO photospheric Carrington maps placed end-to-end. A periodogram is made for each latitude raster time series, and these are stacked by sin(latitude) and plotted horizontally by period All signals shown are significant to at least 99.0% confidence. Some spectral features, lie across the synodic differential rotation curve calculated from day-to-day magnetic field correlations (Komm, Howard, and Harvey, 1993), but others lie completely outside.

*3.2.2 Collective and Independent Modes of Magnetic-Field Rotation*

The purpose of principal components analysis (PCA) is to estimate the minimum number of orthogonal modes that are required to capture the underlying physics of the synoptic magnetic field configurations. PCA (Jackson, 2003; for solar examples see Cadavid, *et al.*, 2005; Cadavid, Lawrence, and Ruzmaikin, 2007; and Lawrence, Cadavid, and Ruzmaikin, 2004) seeks orthogonal eigenmodes of the two-point correlation matrix constructed from a data set. It permits the identification of structures that remain coherent and correlated or that recur throughout a time series. The modes are ranked by their eigenvalues by the degree to which they are correlated to the full data series. In the terminology that we have adopted, the spatial eigenvectors are called "empirical orthogonal functions" or EOFs. Their temporal dependences are the "principal components" or PCs. PCA typically aims to reduce the number of variables required to characterize a physical system.

We begin by choosing data vectors that take advantage of known properties of the magnetic patterns, namely, rapid temporal fluctuation and broad extent in latitude. We thus define a set of (30 × 1) data vectors. One dimension covers the 30 values of sin(latitude) for each synoptic map vertical column and the other dimension, one pixel wide, covers 5º of longitude or about nine hours of time. This maximizes the non-axisymmetry of our analysis. Our concatenated data set includes 28 944 data vectors covering 30 years time. Figure 6 shows the 30 eigenvalues expressed as a percentage of total variance. The two leading eigenvalues are clearly raised above the rest (a secondary drop occurs after 12 eigenvalues). These two leading modes (out of a total of 30) together account for 28.7% of the data variance. The corresponding two leading EOFs are shown as functions of sin(latitude) in Figure 7.

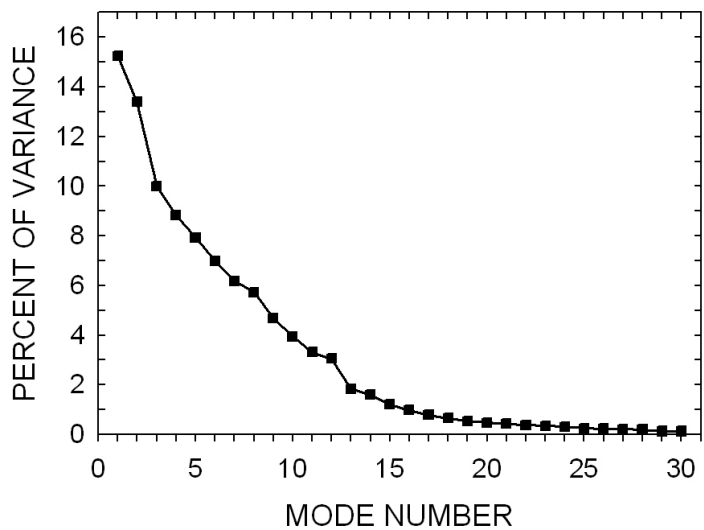

**Figure 6.** The 30 PCA eigenvalues as a percent of variance explained versus mode number.

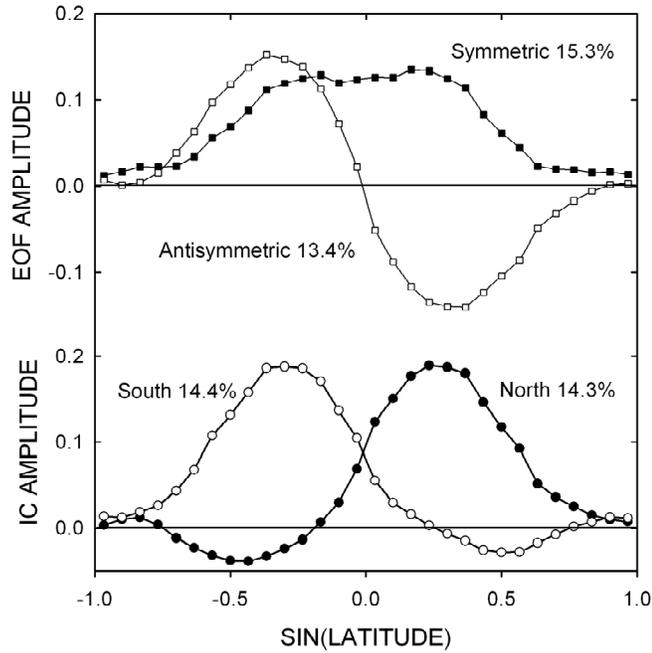

**Figure 7.** Standardized amplitudes of the first two PCA EOFs (squares) and the first two ICA independent components (circles) plotted *versus* sin(latitude) from South (-1) to North (+1). Also shown is the percentage of data variance explained by each mode.

The PCA approach is based on maximal correlation with the full data set subject to orthogonality of the EOFs and PCs. Instead, ICA (Hyvärinen, Karhunen, and Oja, 2001; Stone, 2004; see Cadavid, *et al.*, 2005; Cadavid, Lawrence, and Ruzmaikin, 2007) seeks maximally independent modes and takes into account all order correlations of the data. Our application of ICA consists of recombining the leading EOFs, as established by PCA, to form instead an equal number of Independent Components (ICs). These are fixed by maximizing the degree to which their accompanying time dependences ("mixing vectors" or MVs) are non-Gaussian. This is founded on the central limit idea that when independent time series are combined then the resulting series will be more Gaussian than the originals. Conversely, we say that the least Gaussian variables are most likely to be the physically independent ones.

ICA leads us to the independent components shown in Figure 7. These ICs show North–South mirror symmetry. Each is maximal at |latitude| = 20°. They extend from 15° on one side of the equator to 50° on the other. Although the algorithm used to find these ICs is based on other measures, a quick, quantitative measure of relative Gaussianity is offered by the kurtosis *K* of a signal *Y(t)*:

$K \equiv \langle Y(t)^4 \rangle / \langle Y(t)^2 \rangle^2 - 3$. The kurtosis of a Gaussian time series is exactly 0. For the PCs corresponding to the EOFs of Figure 7 the kurtoses for the Symmetric and Antisymmetric PCs are 1.6 and 1.49, respectively. For the MVs corresponding to the ICs in Figure 7, the kurtoses for N and S are 2.54 and 2.75, respectively – much less Gaussian.

*3.2.3 Characteristic Periods of the Independent Modes*

In Figure 8 we show periodograms for the MVs of the North and South ICs. The northern hemisphere shows dominant peaks at periods 26.45 days and 27.04 days; the southern hemisphere shows a dominant peak at 28.24 days. Numerous weaker periodicities also are significant. As discussed above, the periodogram gives the periodicities that are present over the entire data set. It does not indicate at what time a given periodicity is strong or weak. In Figure 9 we show the Fourier spectral power (see Le Mouël, Shnirman and Blanter, 2007) calculated for the northern and southern MVs at selected periods in a sliding four-year time window. Alternatively, one could apply wavelet methods to this problem (Knaack, Stenflo and Berdyugina, 2005). The Sunspot Number is plotted in Figure 9, along with the spectral power, to permit comparison of the comings and goings of the periodicities to the phase of the solar cycle. Figures 8 and 9 are the chief points of comparison to the results of previous studies. We examine these in in Section 4.

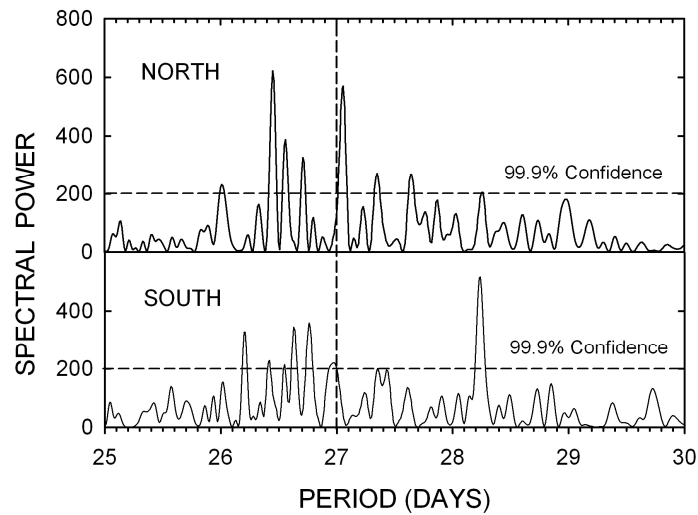

**Figure 8.** Periodograms of the MV time dependences of the North and South ICs in Figure 7 plotted *versus* period in days. The South periodogram has been displaced vertically for clarity.

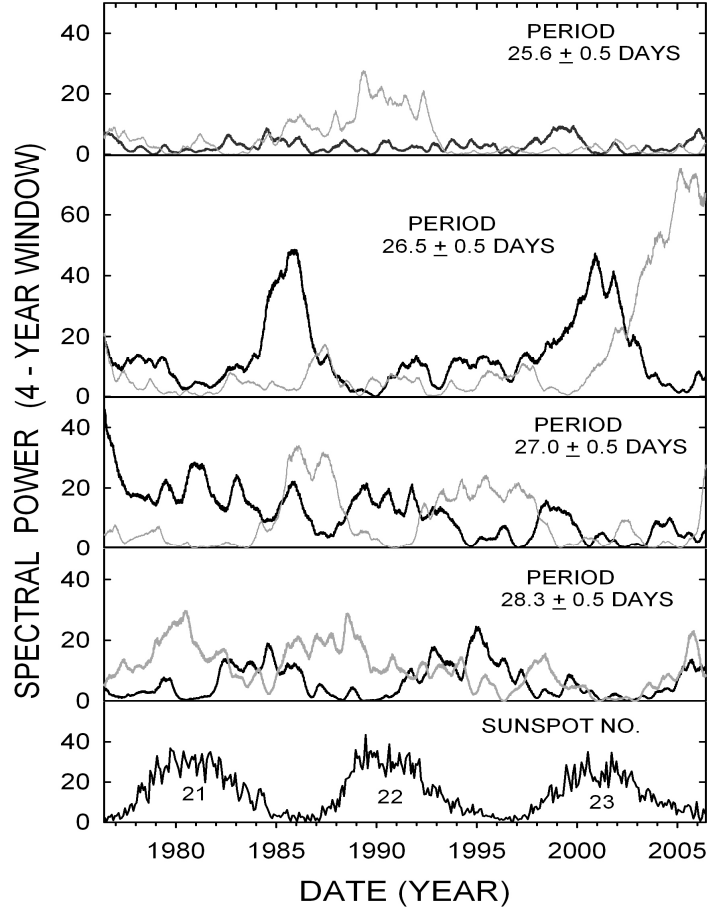

**Figure 9.** Plot of the spectral power in the northern (black) and southern (gray) hemispheric MVs near periods 25.6 ± 0.5 days, 26.5 ± 0.5 days, 27.0 ± 0.5 days and 28.3 ± 0.5 days calculated in a four-year time window. The overall sunspot number is shown for comparison to the activity cycle.

*3.2.4 Antipodal Structures versus Spectral Harmonics*

Fourier or harmonic spectral analyses of time series like those we have investigated in this paper often show the presence of harmonics. That is, in addition to the spectral peaks near 27 days, there will appear peaks near 13.5 days, 9 days, and so forth. It also is sometimes suggested (Bogart, 1982; Knaack, Stenflo and Berdyugina, 2005) that the presence of peaks near 13.5 days points to the favored presence on the Sun of magnetic structures separated by 180º in longitude. However, time series may display strong harmonics in their spectra merely by virtue of not being pure sinusoids. Examples are the IMF polarity reconstruction of Svalgaard (1973), in the form of a square wave, and the unsigned photospheric magnetic field time series that resemble sequences of δ-functions. Spectral peaks near 13.5 days may well have no relation to the presence of antipodal structures.

An alternative way to investigate this question, that avoids the use of harmonic analysis and includes all orders of correlation, is the calculation of "mutual information" (Gallager 1968; see also Lawrence, Cadavid, and Ruzmaikin 1995). We calculate, directly from the data, the joint probability density function $p(Y_i, Y_j, \lambda)$ of finding a value $Y_i$ in a time series and then a value $Y_j$ after a time lag $\lambda$. For *independent* measurements $p(Y_i, Y_j, \lambda) = p(Y_i) p(Y_j)$, the product of the individual probabilities, which also are derivable from the data. The mutual information $I(\lambda)$ is an average over the ensemble of joint measurements: $I(\lambda) = \sum_{i,j} p(Y_i, Y_j, \lambda) \log_2 [p(Y_i, Y_j, \lambda) p(Y_i)^{-1} p(Y_j)^{-1}]$. This is expressed as the negative of an entropy. If the two measurements are independent, then the logarithm vanishes; otherwise $I(\lambda) > 0$. $I(\lambda)$ gives the amount of information, in bits, gained about the value of a number in the time series by knowing the value a time $\lambda$ earlier.

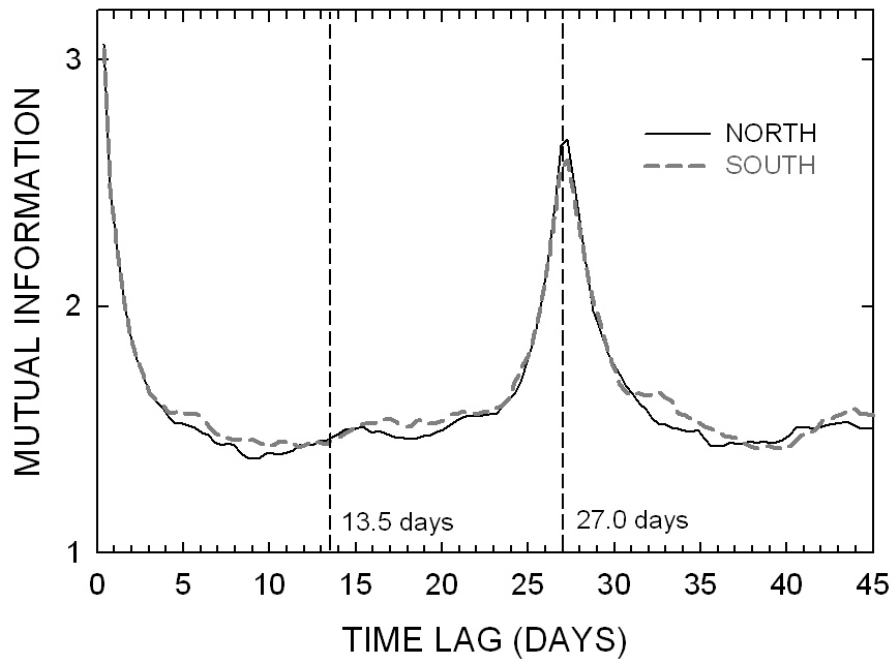

**Figure 10.** The mutual information function averaged separately over the North and South independent component time series plotted versus time lag in days. The vertical dashed lines give references at lags of 27 and 13.5 days.

In Figure 10 we present a calculation of the mutual information for the North and South independent mode MVs. Each of the two series was divided into four parts, and *I(λ)* was calculated for each. These were averaged within each hemisphere. The mutual information shown in Figure 10 is nowhere zero; the presences or absences of magnetic structures at any two times within 45 days of one another are never independent. However, some lags are more dependent than others. *I(λ)* has a strong peak around a lag of 27 days. Thus, knowing about the presence or absence of field now gives strong information about the presence or absence of field 27 days in the future.  (Both correlation and anticorrelation will show as an increase in *I(λ)*; neither are independent.)  Absent, however, is an increase in information at a lag of 13.5 days. The import of this is that if the northern and southern hemispheres of the Sun tend to have preferred longitudes separated by 180º (*e.g.* Berdyugina and Usoskin, 2003) then these are not found in the rigidly rotating modes we have identified. Therefore, the associated "flip-flop" phenomenon, in which the antipodal features exchange dominance, cannot apply to these modes.

## 4. Discussion and Conclusions

The analysis of extended data sets, namely the COHO radial magnetic field from 1963 to 2007 and inferred IMF polarity from 1926 to 2007, supports the finding of Neugebauer, *et al.* (2000) of a dominant periodicity near 27.03 days. This, together with side periodicities, in particular one at 28.25 days, appear in three data sets from three different sources, namely, polar fields from ground-based magnetometer observations, the IMF from *in situ* satellite measurements, and solar coronal force-free extrapolations based on terrestrial observations of the photospheric magnetic field, all from 1976 to 2007. These results are shown in Figures 1 and 2.

Analysis of the 1926 – 2007 IMF polarity data as illustrated in Figure 3 reveals the presence of decadal time spans during which one periodicity or another is dominant in the signal. The periodicity near 27.0 days dominates during 1926 – 1950, 1980 – 1992, and after 2000. The periodicity near 28.3 days dominated from 1950 – 1970. Figure 4 shows that although the amplitude of the dominant periodicity is not constant, it remains generally larger than the others.

Study of the unsigned photospheric magnetic fields, starting with Figure 5, shows the presence of structures (collective modes) that rotate around the Sun at periods near 27 days but that do not trace the differential rotation curve. Instead of showing a rotation period that increases with distance from the equator, they rotate rigidly, with a constant period over a range of latitude from near the Equator in one hemisphere to as far as latitude 35º in the other. The periods seen in the northern and southern hemispheres differ from one another. When we carry out a PCA on a set of maximally non-axisymmetric data vectors, the eigenvalue spectrum (Figure 6) indicates two strong leading modes containing almost 30% of the total variance in the data. The EOFs for these two modes (Figure 7) are one symmetric about the equator and one antisymmetric. However, these EOFs are not strictly unique; ICA recombines the two modes to generate two new modes that are maximally non-Gaussian, and hence maximally independent. The eigenfunctions of these modes, now called ICs, also are shown in Figure 7. These are two mirror symmetric modes, one predominantly in the northern hemisphere, and the other in the southern.

Our two independent modes account for a disproportionate share of the data variance, and they are not perceptibly influenced by surface flows, such as the differential rotation. Further, they are based on synoptic maps which emphasize the emergence of new fields into existing patterns rather than tracking pre-existing flux (Stenflo, 1989). We therefore propose that they are more a direct signature of the solar dynamo than they are of photospheric fields in general. We further conclude that, although the EOFs resemble the standard Legendre polynomials, and therefore may be more mathematically natural, it is the northern and southern hemispheric IC patterns that are the fundamental, physically independent, entities.

Figure 8 shows Lomb-Scargle periodograms of the MVs corresponding to the two ICs. It is no surprise that they differ. Although several periodicities are significant at the 99.9% confidence level, three periods stand out. In the northern mode the strongest peak is at $26.45 \pm 0.03$ days; the second peak is our old friend at $27.04 \pm 0.03$ days. In the southern mode the outstanding period is at $28.24 \pm 0.03$ days. Thus, the $\approx 27.03$ day periodicity serves as a *leitmotiv* that has led us all the way from the fluctuations in the geomagnetic field to the interplanetary magnetic field, to the solar coronal and photospheric fields, and finally to the subsurface dynamo where we have connected it to the northern hemispheric mode. Somewhat less obviously, the $\approx 28.24$ day periodicity has done the same and has been connected to the southern dynamo mode.

Figure 9 indicates the time dependences of the strengths of selected periodicities in the North and South solar modes. Although there is broad agreement, our results differ in details from those of Antonucci, Hoeksema, and Scherrer (1990) and of Knaack, Stenflo, and Berdyugina (2005) due to a different perspective. Because we are interested in the rotation of solar magnetic features, regardless of field strength or activity level, we have divided each magnetic time series by a running standard deviation. This has the purpose of equalizing the signals during activity maxima and minima so as to concentrate on the rotation rates. Thus we are able to see, just for example, enhancement of the $\approx 27.0$ day signal in the South during quiet intervals between cycles 21–22 and cycles 22–23, as well as strong $\approx 26.5$ day signals in the North between cycles 21–22 and in the South after cycle 23.

Svalgaard and Wilcox (1975) reported two basic periodicities in the IMF sector structure: one at 27.0 days and a second at 28.5 days. We have extended this result as shown in Figures 1 through 4. Antonucci, Hoeksema, and Scherrer (1990), using Fourier analysis of WSO and NSO synoptic maps, associated these periods, during solar cycle 21, with solar activity in the northern and southern hemispheres of the Sun, respectively. In our Figure 8 the northern periodicity at $26.45 \pm 0.02$ days and the slightly weaker one at $27.04 \pm 0.02$ days are the two dominant lines in the spectrum of our northern IC mode. Likewise the southern periodicity at $28.24 \pm 0.02$ days dominates the spectrum of the southern IC mode.

Knaack, Stenflo, and Berdyugina (2005) carried out Fourier and wavelet analyses of both signed and unsigned fields in NSO synoptic maps over cycles 21, 22, and 23. They found a dominant $28.1 \pm 0.1$ day periodicity in the South in cycles 21 and 22, as well as a periodicity of 25.0–25.5 days in the South in all three cycles 21, 22, 23. In the North they found a periodicity of $26.9 \pm 0.1$ days during cycle 21, of 28.3–29.0 days during cycle 22 and of $26.4 \pm 0.1$ days during cycle 23.

Comparison to our Figure 9 above shows that the $27.0 \pm 0.5$ day periodicity was strong during the active portions of, not only cycle 21, but also cycle 22. (It also was strong in the southern hemisphere during the minima between cycles 21–22 and 22–23). On the other hand, we found the $28.3 \pm 0.5$ day periodicity to be strong in the southern hemisphere during both cycle 21 and cycle 22 in agreement with Knaack, Stenflo, and Berdyugina (2005). We find that both the 27.0 ± 0.5 day and the 28.3 ± 0.5 day periodicities effectively vanished during cycle 23. However, we found a periodicity at 26.5 ± 0.5 days to be strong in the North during cycle 23, in agreement with Knaack, Stenflo and Berdyugina (2005). These results add substance to the idea that the Sun behaved differently during cycle 23 than in cycles 21 and 22 (de Toma, et al., 2004) and perhaps in earlier cycles not studied by us.

We found the 26.5 ± 0.5 day periodicity to be strong in the North during the minimum between cycles 21–22. Figure 9 indicates that the period at 26.5 ± 0.5 appears to become very strong in the South during the decline of cycle 23. Since Figure 8 shows no strong southern peak at 26.45 ± 0.02 days, we suspect that this represents a general enhancement of the group of weaker southern lines between 26.0 and 26.8 days. We do not see a significant periodicity near 29 days. We do see a periodicity in the South at 25.6 (not shown in Figure 5). However, Figure 9 indicates that this is prominent only during the active part of cycle 22.

Finally, we compare our results to those of Bai (2003), who has collected statistics on the locations on the Sun of flare occurrences during cycles 19 – 23. Bai finds in the South a 28.2 day periodicity in cycle 23, but not in 21 or 22. This is the reverse of our result in which this periodicity appears in cycles 21 and 22 but not 23. In the North a period of 27.0 days is found in cycle 21 only, while we have found it in both cycles 21 and 22. Bai finds a North periodicity of 26.73 days in both cycles 21 and 22 and a North periodicity 27.41 in cycle 21. These are not at dominant peaks in the spectrum of our IC modes in Figure 8, although there are weaker, but significant, lines. All of the periodicities we have listed here, except for the 27.0 day, are listed by Bai (2003) as "double hot-spot" systems. This refers to contemporaneous, co-rotating sources of flaring separated in longitude by about 180º.

We have looked for the reported tendency of solar magnetic structures to appear at favored longitudes 180º apart (Bai, 2003; Berdyugina and Usoskin, 2003). The mutual information approach described above avoids confusion due to the presence of harmonics in Fourier type spectra of the solar signal. We find no sign of the preferred presence of antipodal fields in our two IC components.

# Acknowledgments

The authors thank Drs. Joan Feynman and Leif Svalgaard for helpful conversations during the course of this research. We are pleased to acknowledge and thank those who provided the data used in this work. These included the NASA Coordinated Heliospheric Observations (COHO) data center http://cohoweb.gsfc.nasa.gov/ and the *Wind* and ACE spacecraft teams. Leif Svalgaard made available the IMF polarity reconstructions via the website http://www.leif.org/research/. Extensive use was made of photospheric and source surface synoptic Carrington maps from the Wilcox Solar Observatory at Stanford University (http://wso.stanford.edu/). We thank the Laboratory of Computer and Information Science at Helsinki University of Technology which offers the free FastICA software package in MATLAB (http://www.cis.hut.fi/projects/ica/fastica/).